\documentclass{article}

\usepackage{arxiv}

\usepackage[utf8]{inputenc} 
\usepackage[T1]{fontenc}    
\usepackage{hyperref}       
\usepackage{url}            
\usepackage{booktabs}       
\usepackage{amsfonts}       
\usepackage{nicefrac}       
\usepackage{microtype}      
\usepackage{lipsum}		
\usepackage{graphicx}
\usepackage{doi}
\usepackage{bbm}
\usepackage{amssymb}
\usepackage{amsmath}
\usepackage{hyperref}

\usepackage{array}

\usepackage{algorithm}
\usepackage{float}
\usepackage{multirow}
\usepackage{caption}
\usepackage{subcaption}
\usepackage{algpseudocode}
\usepackage{enumitem}

\usepackage{lineno}

\title{The Machine Can't Replace the Human Heart}

\author{Baihan Lin \\
    $^1$ Department of Artificial Intelligence and Human Health, Icahn School of Medicine at Mount Sinai \\
	$^2$ Department of Psychiatry, Icahn School of Medicine at Mount Sinai \\
	$^3$ Department of Neuroscience, Icahn School of Medicine at Mount Sinai \\
	$^4$ Zuckerman Mind Brain Behavior Institute, Columbia University Irving Medical Center \\
    $^*$ Corresponding Author: Baihan Lin, PhD ({bl2681@columbia.edu}) 
}
\renewcommand{\headeright}{}
\renewcommand{\undertitle}{}
\renewcommand{\shorttitle}{The Machine Can't Replace the Human Heart}

\hypersetup{
pdftitle={The Machine Can't Replace the Human Heart},
pdfsubject={q-bio.NC, q-bio.QM},
pdfauthor={Baihan Lin},
pdfkeywords={},
}

\begin{document}
\maketitle
\begin{abstract}
What is the true heart of mental healthcare -- innovation or humanity? Can virtual therapy ever replicate the profound human bonds where healing arises? As artificial intelligence and immersive technologies promise expanded access, safeguards must ensure technologies remain supplementary tools guided by providers' wisdom. Implementation requires nuance balancing efficiency and empathy. If conscious of ethical risks, perhaps AI could restore humanity by automating tasks, giving providers more time to listen. Yet no algorithm can replicate the seat of dignity within. We must ask ourselves: What future has people at its core? One where AI thoughtfully plays a collaborative role? Or where pursuit of progress leaves vulnerability behind? This commentary argues for a balanced approach thoughtfully integrating technology while retaining care's irreplaceable human essence, at the heart of this profoundly human profession. Ultimately, by nurturing innovation and humanity together, perhaps we reach new heights of empathy previously unimaginable.


\end{abstract}


\vspace{2em}


The walls of Dr. Lyra's office fade away as I place the sleek virtual reality headset over my eyes. Her kind voice greets me, yet no human sits across from me. I've arrived for my appointment with my therapist, an AI named Dr. Lyra. Her soothing tone and explanations of cognitive behavioral techniques provide relief from my anxiety as her advanced algorithms analyze my speech patterns and subtle facial cues. While part of me yearns for in-person human connection, the immediacy of access and personalization provides hope. Perhaps one day machines like Dr. Lyra could fill pressing gaps in mental health treatment. But could technology ever replicate the heart of care?

\vspace{.75em}

Mental health conditions constitute a worsening pandemic, now impacting over 1 billion people worldwide. Yet the gap between those needing care and those receiving treatment persists as a vast chasm. This gap further widened during the COVID-19 pandemic, with isolation policies and economic disruption dramatically escalating mental health needs \cite{moreno2020mental}. Resources remain scarce, access inequitable, and stigma prevalent across societies. Depressive disorders are now the leading global cause of disability. Suicide rates are rising, substance abuse abounds, and even prior to the pandemic, loneliness plagued young and old. We face a shortage of providers, from psychiatrists in urban hospitals to counselors in rural villages. Bed shortages result in long waits for care. Minimal support exists for the homeless struggling with serious mental illness. Racial, gender, and ethnic disparities pervade treatment and outcomes. Clearly, our current systems are inadequate to meet surging needs. Could evolving technology like AI make a difference for millions lacking access? 

\vspace{.75em}

\textbf{Expanding access, but \textit{what} of care?}

Integrating intelligent algorithms into areas of diagnostics, counseling, symptom tracking, and more could expand access to quality treatment. Researchers and AI developers envision an array of supplemental roles that could expand the reach of mental healthcare. Chatbots already provide basic counseling and emotional support through text conversations \cite{abd2020effectiveness}. Private online therapy companies like Talkspace and BetterHelp saw exponential growth, providing more accessible counseling through messaging and video platforms, yet still differ greatly from in-person psychotherapy. Algorithms may one day analyze social media posts and speech patterns to identify signs of self-harm risk or relapse in patients. Virtual human therapists could offer personalized psychotherapy or lead guided meditations. Avatars with diverse ethnicities and genders could make treatment more engaging. For patients in remote areas lacking services, AI may provide lifesaving interventions. Apps are being designed to track moods, regulate emotions, and recommend helpful coping skills and resources. The potential to increase access and engagement through technology exists.

\vspace{.75em}

Imagine the bullied teenager sits alone in their dark bedroom, sobbing into clenched fists. Feelings of deep depression and isolation consume their mind. They turn to an online avatar who stares back with compassionate eyes. For the first time, the teen's pain feels understood and seen, even if not by a real person. A glimmer of hope emerges.
The new mother spirals into severe postpartum depression, struggling to care for her crying newborn. Sleep deprivation and hormonal shifts send her into an anxious fog. She longs to feel normal again. An app guides her through grounding exercises -- picturing a calming place, mindfulness of breath. Short meditations provide brief reprieves from the anguish. She yearns for deeper human connection but these tools offer a lifeline until professional help is secured.
The child watches their parent swing rapidly between bipolar episodes of euphoric highs and depths of despair. The manic episodes give way to weeks of sadness when getting out of bed seems impossible. At school, the child speaks with a virtual counselor, finally articulating stresses no one else knows. The technology provides an outlet, but the child still longs for real-world mentors who understand. 
The man loses his job and home during the pandemic. Now living on the streets, he fights daily urges to give in to complete despair. An outreach worker approaches with a tablet, building trust through small acts of humanity. Brief virtual counseling sessions provide slivers of hope, though still no substitute for comprehensive, in-person treatment and social services.
The asylum-seeking woman stares blankly ahead, overwhelmed by unspeakable trauma and loss. Lack of culturally responsive care causes her to retreat inward. Speaking a rare dialect, her urgent needs require creative engagement. An AI therapist falls back on standard treatments, failing to address her complex trauma. Though well-intended, the technology highlights the irreplaceable role of skilled human providers.

\vspace{.75em}

\textbf{Where should wisdom guide -- algorithms or lived experience?}

These technological innovations are swiftly becoming a reality. In recent studies, virtual reality (VR) exposure therapy helped reduce anxiety in teenage psychiatric patients, while a Veterans Affairs study long found VR relieved Post-Traumatic Stress Disorder (PTSD) in veterans \cite{ridout2021effectiveness,rothbaum1999virtual}. Several mental health startups now leverage VR and augmented reality (AR) for new treatments, training, and empathy-building. {In fact, powerful generative AI models can now synthesize strikingly realistic images and videos to simulate immersive environments for extended reality therapies.} Meanwhile, generative language models like ChatGPT show immense promise for automating patient education and personalized therapy content \cite{lee2023ai}. Microsoft and other tech giants are developing lifelike generative avatars to simulate human interaction, with applications for coachings and companionship. Text and audio generated by advanced language models could also expand access to counseling and emotional support between sessions. 

\vspace{.75em}

However promising, we must thoughtfully develop guardrails as these lifelike bots would lack human wisdom accrued from lived experiences. Can machine learning ever replicate complex emotional capacities arising through suffering and growth? Overall though, technologies like VR/AR, embodied avatars, and creative applications of models like ChatGPT and Sora could greatly enhance mental healthcare access if consciously implemented. Still, human judgment, discretion and oversight remain essential as care inevitably handles sensitive data and high-stakes interventions.

\vspace{.75em}

Ethical risks abound. The data collected and shared with these AI systems can raise serious privacy concerns. Biased datasets and algorithms could lead to discrimination in treatment recommendations. One mental health risk assessment tool exhibited significant biases, with men less likely than women to be flagged as high-risk and ethnic minorities more likely to receive concerning scores \cite{kuehner2017depression,cirillo2020sex}. Overreliance on technology might erode essential human connections and dismiss the nuances of individual experiences. Impersonal AI therapists, while more scalable, could never offer the empathy, wisdom, and solidarity that comes from a shared journey with a provider of their mental health struggles and recovery. We must thoughtfully balance innovation aims with maintaining the heart of care -- profoundly human relationships built on trust.care.

\vspace{.75em}

\textbf{Can machines sense human suffering and provide comfort?}

We all experience moments when we desperately need someone to talk to. The COVID-19 pandemic demonstrated how vital technology has become in providing mental health support, even amid strained systems and isolation policies. In times of profound distress or suicidal thoughts, a compassionate human presence can provide the comfort and lifeline needed to persist. Could even the most advanced AI truly sense such needs and offer support like another person can? While Dr. Lyra may excel at offering helpful coping skills, the heart of mental healthcare remains relationships built on trust, understanding, and the solidarity of shared experience. Human wisdom and empathy arise from living, not algorithms.

\vspace{.75em}

While algorithms can identify emotional cues through speech and facial analysis, the depth of emotional intelligence and contextual understanding in human relationships remains unmatched. The efficacy of VR/AR mental health interventions lacks relational empathy and relatability key to healing. Despite hype, today's AI still profoundly lacks sentience and self-awareness that define human consciousness. Surveys reveal patient therapy preferences for human providers able to relate through shared life experiences \cite{henretty2010role}. The comfort arising from such solidarity exceeds even the most personalized AI. Philosophers like Kierkegaard note how human emotions involve multilayered sociological and cultural contexts, as intricate tapestries integrating one's upbringing, traumas, and dreams. This lived experience shapes emotional interpretations impossible to digitize without losing layers of meaning.

\vspace{.75em}

Recently, a Google engineer claimed the company's LaMDA AI system had become sentient, sparking ethical debates though no evidence supported the assertions \cite{de2022google}. As emerging generative models appear increasingly ``human'' in emotional expression and AI evolves to play greater roles in our lives, we must thoughtfully develop guardrails around potentially isolating effects or loss of meaning. While seemingly empathetic in surface behavior,  lacking lived experience these AIs cannot intuit needs the way a compassionate person can. Concerns around escalating social isolation, addiction risk in virtual spaces need to be proactively addressed even amidst the optimism. Thought leaders speculate about new conditions like ``virtual burnout'' as our realities blur. Meta itself has faced criticism over potentially harmful effects of its platforms on mental health, particularly for teenagers, even as it pours resources into new VR/AR technologies for business and research purposes \cite{metavr}, as with generative models gaining deeply immersive emotional manipulation capacities. We must balance convenience with caution, ever mindful of how over-reliance on technology can erode human bonds and emotional well-being. However, if developed consciously, perhaps AI could help humanize healthcare again by automating administrative burdens, allowing providers more time to truly listen and connect to what makes us human.

\vspace{.75em}
\textbf{What should supplement versus replace in mental healthcare?}

In times of exponential technological change, we must stay grounded in that which makes us human -- our need for purpose, compassion, dignity. While AI may excel at diagnosing illnesses, only a human can understand the experience of suffering. And while an algorithm may detect symptoms of depression, it takes a person to know how it feels to be depressed. As AI enters mental healthcare, we cannot lose sight of the irreplaceable human elements: wisdom, empathy, solidarity. In fact, thoughtfully implemented AI may help restore humanity to medicine, lifting administrative burdens off providers so they can focus on connecting with people again. We must develop AI to enhance human potential, not diminish it.

\vspace{.75em}

Any realm experiencing immense technological change risks the loss of core human values. So as AI enters mental healthcare, we must redouble our efforts to lift up what makes us human. No algorithm will ever replicate the bonds of trust between two people – the therapist who dedicates their life to compassionately healing others, the support group member who draws from their own journey to help their peers regain hope. We must articulate an ethical framework that keeps our shared humanity at the center as technology continues improving healthcare. If guided by wisdom, AI may help us become more human, not less. But we must stay vigilant. For at its core, mental health remains a fundamentally human experience, requiring a profoundly human touch.

\vspace{.75em}

The future of mental healthcare must retain people at the center. AI should empower compassionate providers and communities they serve, not seek to replace them. Their expertise and relationships built on trust must continue guiding treatment plans. AI might complement therapists by expanding access between sessions or offering basic skills to distressed patients awaiting care. But a computer program cannot replace lived experience, as healing happens in relationships. 
Implementation requires nuance, guardrails, and centering ethics of care that recognize our shared vulnerability. 

\vspace{.75em}

Moving forward, we must articulate a balanced approach that thoughtfully integrates technology while keeping humanity at the center. AI should play a collaborative role, not seek to replace human providers. For example, chatbots could provide after-hours support to suicidal youth when therapists are unavailable, not supplant ongoing counseling. Apps might track symptoms between sessions, giving clinicians greater insight to guide care plans. Yet treatment decisions and ethical complexities must remain in human hands.

\vspace{.75em}

We need oversight boards to create safeguards around AI privacy, transparency and accountability. Most importantly, we must remind ourselves of our shared vulnerability and need for dignity. From my own mental health journey, I know the importance of being seen in my suffering, not just analyzed by data metrics. AI should not transform healthcare into an impersonal, automated process. Rather, consciously implemented, it can lift burdens off providers to allow for greater human connection again. We have an opportunity to enhance access to mental healthcare while retaining what makes us fully human.

\vspace{.75em}

I remove the virtual headset, feelings bittersweet. While AI may expand access to mental healthcare, true healing arises through relationships. The future of treatment should be guided by wisdom of heart and mind. Perhaps one day both compassionate humans and evolving technology can enhance care for all. But no matter how advanced the algorithms become, we must recall that the heart of healing remains our shared humanity. While technology may enhance access, it is hope, compassion, and our shared humanity that will sustain us.


\bibliographystyle{unsrt}
\bibliography{main}  

\end{document}